\documentclass{article}
\usepackage[T1]{fontenc}
\usepackage[utf8]{inputenc}
\usepackage{ismir}
\usepackage{amsmath,url}

\usepackage{graphicx}
\usepackage[square,numbers,sort&compress]{natbib}

\usepackage{amssymb}
\usepackage{booktabs}       
\usepackage{amsfonts}       
\usepackage{nicefrac}       
\usepackage{microtype}      
\usepackage[dvipsnames]{xcolor}         
\usepackage{fontawesome5}
\usepackage{multicol}
\usepackage{multirow}
\usepackage{tikz}
\usepackage{pifont}
\usetikzlibrary{positioning}
\usepackage{enumitem}
\makeatletter
\@namedef{ver@setspace.sty}{2011/12/19} 
\makeatother
\usepackage{musicography}

\newcommand{\tabitem}{~\llap{\textbullet}~}

\definecolor{cgood}{HTML}{1f782c}   
\definecolor{cbad}{HTML}{9e0e0e}    

\newcommand{\fatcheckmark}{\ding{52}}
\newcommand{\fatcross}{\ding{56}}

\title{\textsc{M\MakeLowercase{idi}-LLM}: 
Improving Text-to-MIDI Music Generation\\via Adapting Large Language Models
}

\multauthor
  {Shih-Lun Wu$^{\,\musSharp}$ \hspace{1.1cm} Dave Carlton$^{\,\musFlat}$ \hspace{1.1cm} Ryan Miyakawa$^{\,\musFlat}$ }
  {{\bf Yoon Kim$^{\,\musSharp}$ \hspace{5mm} Chris Donahue$^{\,\musNatural\raisebox{0.1ex}{\musEighth}}$ \hspace{5mm} Cheng-Zhi Anna Huang$^{\,\musSharp\raisebox{0.1ex}{\musEighth}}$}\\
  $^{\,\musSharp}$ Massachusetts Institute of Technology \hspace{6mm} $^{\,\musFlat}$ Hooktheory \hspace{6mm}$^{\,\musNatural}$ Carnegie Mellon University \\
  {\tt\small \{slseanwu, huangcza\}@mit.edu} \hspace{1cm} {\small {\scriptsize ${\raisebox{0.1ex}{\musEighth}}\,$} Equal senior contribution}
  \vspace{-2ex}}

\def\authorname{S.-L. Wu et al}
\usepackage[bookmarks=false,pdfauthor={\authorname},pdfsubject={\pdfsubject},hidelinks]{hyperref}
\hypersetup{                    
  colorlinks,
  linkcolor={OliveGreen},
  citecolor={Blue},
  urlcolor={BlueViolet}
}

\begin{document}

\maketitle


\begin{abstract}
\vspace{-2mm}
We present \textsc{Midi-LLM}, a recipe that improves multitrack text-to-MIDI generation via adapting Large Language Models (LLMs).
\textsc{Midi-LLM} expands an LLM's text vocabulary to include MIDI tokens,
and employs a two-stage training pipeline:
(i)~unimodal continued pretraining on music-adjacent text and standalone MIDIs, and
(ii)~multimodal supervised finetuning on text--MIDI pairs.
Our instantiation of \textsc{Midi-LLM} based on Llama 3.2 (1B)
outperforms the recent Text2midi model in both text control and musical quality,
and readily integrates with optimized inference ecosystems like \texttt{vLLM}.
To align with real-world songwriting workflows,
we further finetune our \textsc{Midi-LLM} on the TheoryTab dataset for text-conditioned lead sheet (i.e., melody + chords) generation and infilling.
A comprehensive ablation study validates the synergy between LLM text pretraining, standalone MIDI pretraining, and supervised text-to-MIDI finetuning.
Finally, 
an \textit{in-the-wild} blind user study conducted in a \emph{real-world} creative workflow at scale with 58 Hookpad Aria users and 4,002 generated outputs
demonstrates that our \textsc{Midi-LLM} achieves the highest acceptance rate
in zero-to-one lead sheet generation
over baselines without text control or LLM pretraining,
confirming its efficacy in human-AI music co-creation.


\end{abstract}

\vspace{-2mm}

\section{Introduction}
Recent advances in text-to-music models empowered users to generate realistic-sounding audio music from simple natural language prompts~\citep{copet2023simple,evans2024fast}.  
However, even with fine-grained control mechanisms~\citep{wu2024music,tsai2025musecontrollite} 
and ongoing efforts in generative audio editing~\citep{zhang2024musicmagus,tsai2024audio,wu2026stemphonic},
audio waveform outputs remain difficult to precisely edit, restructure, or reuse in downstream workflows.  
This constrains their ability to support the kind of iterative human-AI collaboration that Large Language Models (LLMs) have enabled in text domains.  
In contrast, symbolic-domain models most commonly generate MIDI outputs~\citep{huang2019music,wu2023compose,thickstun2024anticipatory}
that 
allow direct editing and reuse. 
Such models have been praised by musicians for promoting creative agency~\citep{donahue2025hookpad,kim2025amuse}, and have seen integrations into digital music software~\cite{malandro2023composer,malandro2024composer}.  
Yet, symbolic models largely lack effective free-form text control typically found in audio-domain counterparts, limiting their utility for rapid initial ideation and/or instruction-based editing.
They also often rely on custom model architectures~\citep{lu2023musecoco,bhandari2025text2midi,wang2025notagen} that may each require bespoke engineering efforts to accelerate inference for production-scale deployment.

\begin{figure*}
  \centering
  \includegraphics[width=0.96\textwidth]{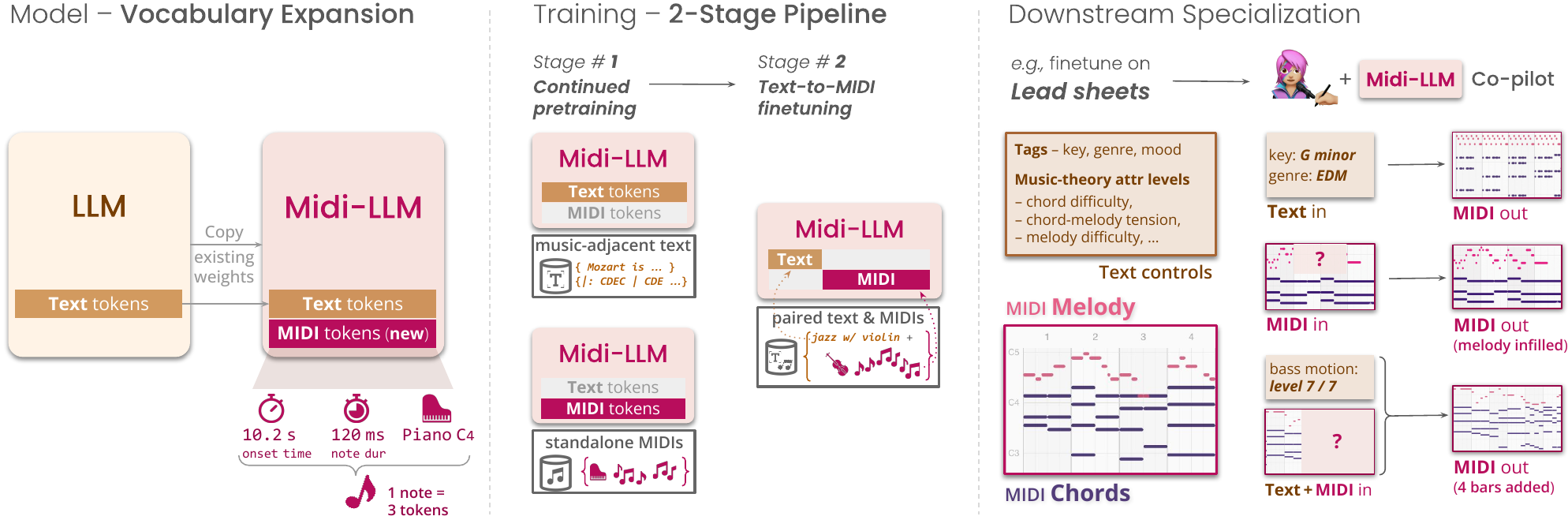}
  \vspace{-4mm}
  \caption{\textsc{Midi-LLM} recipe overview. 
  We instantiate our \textsc{Midi-LLM} by expanding the token embedding layer of \textit{Llama 3.2} (1B)~\citep{grattafiori2024llama} with the \textit{Anticipatory Music Transformer (AMT)}~\citep{thickstun2024anticipatory} MIDI vocabulary.
  We then train the LLM in two stages to achieve text-to-MIDI generation and infilling.
  The resulting model can further be finetuned for downstream MIDI generation tasks involving any combination of text and MIDI conditions, which we demonstrate on TheoryTab~\citep{donahue2022melody} lead sheets.
}\label{fig:overview}
\end{figure*}

To imbue symbolic models with effective text control
and to leverage continuous advancements in the quality and efficiency of open-weight LLMs, 
we propose \textsc{Midi-LLM}, a recipe to adapt LLMs for text-to-MIDI generation and editing.  
Our motivation is twofold:
(i)~LLMs encode broad world knowledge and emotional semantics, making them well-suited for conditioning on text prompts, and
(ii)~a monolithic LLM architecture fits into an ecosystem with well-developed, easy-to-use inference-time optimizations.
To validate our recipe, we explore a first instantiation by expanding
the token embedding layer of the Llama 3.2 (1B) LLM~\citep{grattafiori2024llama} to incorporate MIDI tokens (Sec.~\ref{subsec:expand}) based on the Anticipatory Music Transformer~\citep{thickstun2024anticipatory} vocabulary 
and framework built with MIDI generation and infilling support.  
We then train the model in two stages totaling 4B+ tokens: first on music-adjacent text and standalone MIDIs as \textit{unimodal} continued pre-training, and then on multitrack Lakh MIDI~\citep{raffel2016lmd} data paired with text captions~\citep{melechovsky2024midicaps} as \textit{multimodal} supervised finetuning (Sec.~\ref{subsec:training}).
Since we preserve the original LLM's parameter structure, we can easily leverage \texttt{vLLM}~\citep{kwon2023efficient,shaw2024llm} for accelerated inference.
We perform a quantitative 
comparison with the recent Text2midi~\citep{bhandari2025text2midi} model (Table~\ref{tab:vs-text2midi}), showing that our model achieves higher quality, stronger text control, and 7--13$\times$ faster inference.

After our two-stage training, the resulting model effectively serves as a MIDI foundation model that can be further finetuned and specialized for any downstream MIDI generation tasks with text and/or MIDI conditioning.
We demonstrate this capability by further finetuning on TheoryTab~\citep{donahue2022melody} lead sheets, i.e., compositions containing a melody and a chord progression, paired with textual mood, genre, and music-theoretical labels.
We focus on 
this downstream task as recent co-pilot systems~\citep{donahue2025hookpad,kim2025amuse} helping with these core elements of harmonic planning and melodic sketching have provided high practical value to songwriters by fostering a human-AI co-creation workflow.
With this relatively small-scale (67K pieces, 34M tokens) finetuning dataset, we conduct an ablation study of prior \textsc{Midi-LLM} training stages (Table~\ref{tab:ablations}),
validating the synergy of LLM text pretraining, unimodal text/MIDI continued pretraining, and supervised text-to-MIDI training,
both in terms of validation LM loss and symbolic music metrics evaluating quality and control adherence.
Furthermore, we implement classifier-free guidance (CFG)~\citep{copet2023simple,ho2022classifier} on text prompts,
offering a tunable parameter at inference to trade-off quality for stronger text control, and to balance between conditioning from surrounding MIDI context 
and user intent provided in the text prompt
in infilling scenarios (Fig.~\ref{fig:cfg-tradeoff}).

Finally, we conduct an in-the-wild lead sheet co-creation user study (Sec.~\ref{sec:user-study}) with 58 Hookpad Aria~\citep{donahue2025hookpad} subscribers.
Our study has an interface where users can write and edit the melody and chord tracks, and specify genre, mood, and music-theoretical tags to get from-scratch generations or span infill suggestions from a co-pilot model.
Co-pilot requests are blindly and randomly routed to our TheoryTab-finetuned \textsc{Midi-LLM}, or two other baselines: one with stripped text prompt, and the other a MIDI-only model without text conditioning.
Acceptance rates collected from a total of 4,002 model generations substantiate the real-world advantage of \textsc{Midi-LLM} in zero-to-one music ideation.

Our contributions can be summarized as follows:
\begin{itemize}[nolistsep,itemsep=1pt,leftmargin=12pt]
    \item We provide a recipe for adapting LLMs for text-to-MIDI generation and infilling by vocabulary expansion and subsequent two-stage training. This recipe readily integrates with optimized LLM serving ecosystems.
    \item We present a concrete instantiation of \textsc{Midi-LLM} based on Llama 3.2 (1B), showing that our model outperforms the recent Text2midi model.
    \item We demonstrate that our \textsc{Midi-LLM} instantiation can be specialized for the downstream workflow of lead sheet composition, and conduct an ablation study to showcase the synergy between prior large-scale (1B+ tokens) text, MIDI, and text-to-MIDI training stages on small-scale ($<$50M tokens) TheoryTab downstream data.
    \item We deploy an in-the-wild user study in an online human-AI lead sheet co-creation setting. Acceptance rates from a total of 58 participants and 4K+ co-pilot model generations confirm the real-world value of our \textsc{Midi-LLM} recipe in zero-to-one music ideation.
\end{itemize}
We encourage readers to explore our
live demo,\footnote{Live demo \& sound examples: \url{https://midi-llm-demo.vercel.app}.}\,model checkpoint,\footnote{Model: \url{https://huggingface.co/slseanwu/MIDI-LLM_Llama-3.2-1B}.}\,and inference code.\footnote{Inference code: \url{https://github.com/slSeanWU/MIDI-LLM}.}



\section{Background: MIDI Tokenization}\label{sec:token}
Various approaches have been proposed to tokenize MIDI music for language modeling.
Popular families include:
(i)~\textit{MIDI-like} tokens~\citep{oore2018time,huang2019music},
(ii)~metered Revamped MIDI tokens (\textit{REMI})~\citep{huang2020pop,hsiao2021compound,wu2023compose}, and
(iii)~text-based \textit{ABC}-derived notations~\citep{yuan2024chatmusician,qu2024mupt,wang2025notagen}.
For \textsc{Midi-LLM}, we adopt the MIDI-like tokenization from Anticipatory Music Transformer (AMT)~\citep{thickstun2024anticipatory}.
The AMT{\small (arrival-time)} vocabulary represents each musical note as a token triplet:
\begin{itemize}[nolistsep,itemsep=1pt,leftmargin=12pt]
    \item \textbf{Onset time:} $0^{\text{th}}$ to $100^{\text{th}}$ second with $10$\,ms quantization.
    \item \textbf{Duration:} $0$ to $10$ seconds with $10$\,ms quantization.
    \item \textbf{Instrument-pitch} joint token\textbf{:} $129$ inst.~$\times$ $128$ pitches.
\end{itemize}
This amounts to 27.5K possible tokens.
Additionally,
AMT uses a mirroring set of \textit{anticipated} tokens to represent future notes (i.e., right context) given to the LM as conditions for music infilling tasks, hence a total vocabulary size of 55K.
An anticipated note (i.e., token triplet) at time $\mathsf{T + \Delta}$
is \textit{interleaved} into the sequence once the (regular) notes progress to time $\mathsf{T}$.
Following~\citep{thickstun2024anticipatory}, we set {$\mathsf{\Delta} := 5$} seconds.

Compared to REMI-based tokens and ABC-derived notations, AMT and similar MIDI-like vocabularies are more widely applicable as they do not require metered (i.e., beat-synchronized) music data.
Also, they can be more expressive than ABC notations as MIDI natively encodes details like micro-timing and per-note dynamics~\citep{oore2018time, huang2019music}.



\begin{table}
\scriptsize
\setlength{\tabcolsep}{1.25pt}
\centering
\begin{tabular}{lp{3.35cm}p{1.8cm}p{1.8cm}}
\toprule
& \multicolumn{2}{c}{\footnotesize\textbf{\textit{Continued Pretraining}}} & \multicolumn{1}{c}{\footnotesize\textbf{\textit{Finetuning}}} \\
\cmidrule(lr){2-3} \cmidrule(lr){4-4}
&
\multicolumn{1}{c}{\tikz[baseline=(current bounding box.base)] \node[rounded corners=3pt, fill=Bittersweet!90!black, inner sep=3pt, outer sep=0pt](text){{\textcolor{white}{\textbf{Text}}}};}
& \multicolumn{1}{c}{\tikz[baseline=(current bounding box.base)]\node[rounded corners=3pt, fill=OrangeRed!90!black, inner sep=3pt, outer sep=0pt](midi){{\textcolor{white}{\textbf{MIDI}}}};}
& \multicolumn{1}{c}{\tikz[baseline=(current bounding box.base)]{
\node[rounded corners=3pt, fill=Bittersweet!90!black, inner sep=3pt, outer sep=0pt] (text2) {\textcolor{white}{\textbf{Text}}};
\node[right=0pt of text2, inner sep=0pt, outer sep=0pt] (to) {\textbf{-to-}};
\node[rounded corners=3pt, fill=OrangeRed!90!black, inner sep=3pt, outer sep=0pt, right=0pt of to] (midi2) {\textcolor{white}{\textbf{MIDI}}};
}} \\
\midrule
\multirow{1}{*}[-4ex]{\textbf{Dom.}} & 
\begin{tabular}[t]{@{}l@{}}
\tabitem Music web articles {\tiny (56\%)} \\
\tabitem Music knowledge \tiny{(GPT-4)} {\tiny (17\%)} \\
\tabitem QAs: ABC-notated music {\tiny (16\%)} \\
\tabitem QAs: common knowledge {\tiny (12\%)}
\end{tabular} & 
\multirow{1}{*}[-2.2ex]{\begin{tabular}[t]{@{}l@{}}
\textit{Standalone} \\
multitrack MIDIs
\end{tabular}
}
& 
\multirow{1}{*}[-2.2ex]{\begin{tabular}[t]{@{}l@{}}
Multitrack MIDIs \\
\textit{paired w/ text}
\end{tabular}
}\\
\midrule
\multirow{1}{*}[-1.2ex]{\textbf{Dset.}} & 
\multirow{1}{*}[-1.2ex]{\begin{tabular}[t]{@{}l@{}}
(Subset of) MusicPile
\citep{yuan2024chatmusician}
\end{tabular}
} & 
\multirow{1}{*}[-1.2ex]{\begin{tabular}[t]{@{}l@{}}
GigaMIDI \citep{lee2025gigamidi}
\end{tabular}
}& 
\begin{tabular}[t]{@{}l@{}}
MidiCaps \scriptsize{(text)}
\citep{melechovsky2024midicaps} \\
+ Lakh \scriptsize{(MIDI)} 
\citep{raffel2016lmd}
\end{tabular}
\\
\midrule
\textbf{Size} & \footnotesize 1.69\,B tokens & \footnotesize 1.38\,B tokens & \footnotesize 1.71\,B tokens \\
\midrule
\multirow{1}{*}[-1ex]{\textbf{Seqlen}} & \footnotesize \multirow{1}{*}[-0.7ex]{2048 } & \footnotesize \multirow{1}{*}[-0.7ex]{2048 {\scriptsize($\approx$30\,sec.)}} & 
\footnotesize
\begin{tabular}[t]{@{}l@{}}
Text \;\, $\leq$256 \\
MIDI \; 2048
\end{tabular} \\
\bottomrule
\end{tabular}
\vspace{-2mm}
\caption{\textsc{Midi-LLM} two-stage training data summary.}
\label{tab:data}
\end{table} 

\section{Method}
\subsection{LLM Vocabulary Expansion}\label{subsec:expand}
To incorporate the AMT~\citep{thickstun2024anticipatory} music tokens (cf.~Sec.~\ref{sec:token}) into an LLM, a na\"ive approach is to serialize them as a text string (e.g., {\small `\textit{<onset 10.2s> <duration 120ms> <piano, C4>}'}) and feed to the LLM's text tokenizer.
However, this would unnecessarily lengthen the sequence and increase compute overhead.
Therefore, 
we expand the LLM's token embedding weights, {\textcolor{Bittersweet!90!black}{ $\mathbf{E}_{\text{LLM}}$}}, via:
\begin{equation}
    \mathbf{E}_{\textsc{Midi-LLM}} := \left[ \textcolor{Bittersweet!90!black}{\mathbf{E}_{\text{LLM}}^{\top}} \;\; \textcolor{OrangeRed!90!black}{\mathbf{E}_{\text{AMT}}^{\top}} \right]^{\top} \in \mathbb{R}^{\left(|\mathcal{V}_{\text{LLM}}| + |\mathcal{V}_{\text{AMT}}|\right) \times D} \, ,
    \label{eq:vocab-expand}
\end{equation}
where $|\mathcal{V}_{\text{LLM}}|$ is the original text vocabulary size,
$|\mathcal{V}_{\text{AMT}}|$ is the new music vocabulary size (55K), and $D$ is the model's hidden-state dimension.
With this design, each note becomes exactly three tokens to the LLM.
We randomly initialize $\textcolor{OrangeRed!90!black}{\mathbf{E}_{\text{AMT}}^{\top}}$ and the corresponding LM head (i.e., unembedding layer) part, and train the entire LLM subsequently.

\subsection{Two-Stage Training}\label{subsec:training}
\textsc{Midi-LLM}
employs a two-stage training pipeline:
(i)~\textit{unimodal} continued pretraining on broad  text and MIDI data,
and then (ii)~\textit{multimodal} supervised finetuning on MIDIs paired with text captions.
The standard next-token prediction objective is used in both stages.
A summary of datasets used in \textsc{Midi-LLM} training is shown in Table~\ref{tab:data}.

\subsubsection{Continued Pretraining (CPT)}
Following common practices to adapt LLMs for certain domains~\citep{gururangan2020dont,ibrahim2024simple}, we perform unimodal continued pretraining with two primary goals:
(i)~to surface the musical knowledge (that exists in text) the LLM might have seen in its initial pretraining, and
(ii)~to teach it the syntax and structure of MIDI data under AMT's framework~\citep{thickstun2024anticipatory}.
We compile a dataset of around 3B tokens for this stage:
\begin{itemize}[topsep=1pt,itemsep=0pt,leftmargin=12pt]
    \item \textbf{Music-adjacent text:} A subset of \textit{MusicPile}~\citep{yuan2024chatmusician} containing primarily music-related articles, GPT-4 synthetic music knowledge, and ABC-notated music.
    \item \textbf{Standalone MIDIs:} The multitrack music pieces in \textit{GigaMIDI}~\citep{lee2025gigamidi}, excluding overlapping examples with Lakh MIDI~\citep{raffel2016lmd} reserved for the later stage.
\end{itemize}
We mix text and MIDI examples in one training batch,
and augment the MIDI token sequences on-the-fly for anticipation (i.e., infilling) leveraging the AMT framework.

\subsubsection{Text-to-MIDI Supervised Finetuning (SFT)}
To enable music generation from textual descriptions, 
we finetune the model on paired (text, MIDI) data, 
teaching it to translate musical concepts expressed in text into MIDI notes.
We construct each example with a text prompt from \textit{MidiCaps}~\citep{melechovsky2024midicaps} as the \textit{instruction prefix}, which contains attributes like genre, mood, instrumentation, tempo, key, and chords of the music,
followed by the AMT tokens of the corresponding MIDI in \textit{Lakh\,MIDI~(LMD)}~\citep{raffel2016lmd}.
For augmentation, we create span infilling examples with the notes of either \textit{all} or a \textit{subset} of active instruments as the infill target,
and utilize \textit{Qwen2.5-Omni}~\citep{xu2025qwen2}, an LLM capable of music captioning,
to produce text prompts for these target infill segments.
The token count for this stage is around 1.7B pre-augmentation, and 5.1B (tripled) post-augmentation.
The resulting model may serve as a versatile backbone amenable to downstream specialization
for any MIDI generation tasks involving arbitrary combinations of textual and musical conditioning.

\begin{table}
\small
\renewcommand{\arraystretch}{1.3}
\setlength{\tabcolsep}{3pt}
\centering
\begin{tabular}{lcrccrr}
\toprule
\multicolumn{3}{c}{} & \textbf{Qlty.} & \textbf{Ctrl.} & \multicolumn{2}{c}{\textbf{Inference RTF} $\uparrow$} \\
 \cmidrule(lr){4-5} \cmidrule(lr){6-7}
\textbf{Model} & \textbf{Size} & \textbf{Prec.} & \textit{FAD}\,$\downarrow$ & \textit{CLAP}\,$\uparrow$ & bsz$=$1 & bsz$=$4  \\
\midrule
Text2midi & 0.3B & fp32 & 0.818 & 0.187 & 0.56 & 1.06 \\ \hline
\multirow{2}{*}[-0.5ex]{\renewcommand{\arraystretch}{0.9}\begin{tabular}{@{}l@{}}\textsc{\textbf{Midi-LLM}}\\\textbf{(Ours)}\end{tabular}} & \multirow{2}{*}[-0.5ex]{1.5B} & bf16 & \textbf{0.173} & \textbf{0.221} & 3.33 & 11.48 \\
& & fp8 & 0.216 & 0.218 & \textbf{4.02} & \textbf{14.17} \\
\bottomrule
\end{tabular}
\vspace{-3mm}
\caption{Evaluation on text-to-multitrack MIDI generation (MidiCaps + LMD).
\textsc{Midi-LLM} accomplishes better outputs and faster generation than Text2midi~\citep{bhandari2025text2midi}.
Real-time factor (RTF) is measured with 2K seqlen on an L40S GPU.
}
\label{tab:vs-text2midi}
\end{table} 

\begin{table*}
   \small
   \setlength{\tabcolsep}{6.5pt}
\renewcommand{\arraystretch}{1.08}
   \centering
    \begin{tabular}{lcccrrrrrrr}
    \toprule
     & \multicolumn{3}{c}{\textbf{Prior Training} (\# tokens)} & \textbf{LM Loss} & \multicolumn{6}{c}{\textbf{Music Quality}} \\
    \cmidrule(lr){2-4}\cmidrule(lr){5-5}\cmidrule(lr){6-11}
     & \multirow{2}{*}[0.1ex]{\renewcommand{\arraystretch}{0.9}\begin{tabular}{@{}c@{}}\footnotesize Llama\,3\\\textbf{Text}\\\footnotesize{(9T)}\end{tabular}}
     & \multirow{2}{*}[0.1ex]{\renewcommand{\arraystretch}{0.9}\begin{tabular}{@{}c@{}}\footnotesize Giga-\\\textbf{MIDI}\\\footnotesize{(1.4B)}\end{tabular}}
     & \multirow{2}{*}[0.1ex]{\renewcommand{\arraystretch}{0.9}\begin{tabular}{@{}c@{}}\footnotesize Lakh\\\textbf{Text--MIDI}\\\footnotesize{(1.7B)}\end{tabular}}
     & \multirow{2}{*}[-0.5ex]{\shortstack[c]{Valid.\\\textit{NLL} $\downarrow$}} & \multicolumn{3}{c}{\textbf{Chroma} \, \textit{JS Distance} $\downarrow$} & \multicolumn{3}{c}{\textbf{Groove} \, \textit{JS Distance} $\downarrow$} \\
    \cmidrule(lr){6-8}\cmidrule(lr){9-11}
    \textbf{Ablations} & & & & & Melody & Chords & Both & Melody & Chords & Both \\
    \midrule
    \#\textbf{1} -- No PT & {\color{cbad}\fatcross} & {\color{cbad}\fatcross} & {\color{cbad}\fatcross} & 0.543 & 0.066 & 0.103 & 0.091 & 0.263 & 0.295 & 0.245 \\
    \#\textbf{2} -- Text PT & {\color{cgood}\fatcheckmark} & {\color{cbad}\fatcross} & {\color{cbad}\fatcross} & 0.471 & 0.079 & 0.109 & 0.087 & 0.245 & 0.228 & 0.246 \\
    \#\textbf{3} -- MIDI PT & {\color{cbad}\fatcross} & {\color{cgood}\fatcheckmark} & {\color{cbad}\fatcross} & 0.440 & \underline{0.059} & 0.082 & 0.078 & 0.202 & 0.194 & 0.205 \\
    \#\textbf{4} -- Text\,\&\,MIDI PT & {\color{cgood}\fatcheckmark}$^{\ast}$ & {\color{cgood}\fatcheckmark} & {\color{cbad}\fatcross} & 0.443 & 0.069 & \textbf{0.077} & \textbf{0.061} & \textbf{0.163} & \underline{0.156} & \underline{0.166} \\
    \#\textbf{5} -- Text--MIDI SFT  & {\color{cbad}\fatcross} & {\color{cbad}\fatcross} & {\color{cgood}\fatcheckmark} & \underline{0.428} & 0.068 & 0.090 & 0.081 & 0.222 & 0.192 & 0.229 \\
    \hline
    \textbf{\textsc{Midi-LLM} (Ours)} & {\color{cgood}\fatcheckmark}$^{\ast}$ & {\color{cgood}\fatcheckmark} & {\color{cgood}\fatcheckmark} & \textbf{0.423} & \textbf{0.056} & \underline{0.078} & \underline{0.066} & \underline{0.170} & \textbf{0.148} & \textbf{0.159} \\
    \bottomrule
    \end{tabular} \\

\vspace{1.5mm}
\begin{tabular}{lcccrr@{\hskip 12pt}rrrrr}
    \toprule
     & \multicolumn{3}{c}{\textbf{Prior Training} (\# tokens)} & \multicolumn{7}{c}{\textbf{ Textual Controllability}} \\
    \cmidrule(lr){2-4}\cmidrule(lr){5-11}
     & \multirow{2}{*}[0.1ex]{\renewcommand{\arraystretch}{0.9}\begin{tabular}{@{}c@{}}\footnotesize Llama\,3\\\textbf{Text}\\\footnotesize{(9T)}\end{tabular}}
     & \multirow{2}{*}[0.1ex]{\renewcommand{\arraystretch}{0.9}\begin{tabular}{@{}c@{}}\footnotesize Giga-\\\textbf{MIDI}\\\footnotesize{(1.4B)}\end{tabular}}
     & \multirow{2}{*}[0.1ex]{\renewcommand{\arraystretch}{0.9}\begin{tabular}{@{}c@{}}\footnotesize Lakh\\\textbf{Text--MIDI}\\\footnotesize{(1.7B)}\end{tabular}}
     & \multicolumn{2}{c}{\textit{Clf.~Score} {$\mathbf{\Delta}$} $\uparrow$} & \multicolumn{5}{c}{\shortstack[c]{\textbf{Music-Theoretical Attr.~Level} \, \textit{Spearman's} $\mathbf{\rho} \uparrow$}} \\
    \cmidrule(lr){5-6}\cmidrule(lr){7-11}
    \textbf{Ablations} & & & & \textbf{Genre} & \textbf{Mood} & \multicolumn{1}{c}{\shortstack[c]{Ch-D}} & \multicolumn{1}{c}{\shortstack[c]{Ch-N}} & \multicolumn{1}{c}{\shortstack[c]{Mel-D}} & \multicolumn{1}{c}{\shortstack[c]{Ch-Mel-T}} & \multicolumn{1}{c}{\shortstack[c]{Bass-M}} \\
    \midrule
    \#\textbf{1} -- No PT & {\color{cbad}\fatcross} & {\color{cbad}\fatcross} & {\color{cbad}\fatcross} & 0.203 & 0.177 & 0.593 & 0.617 & 0.213 & 0.211 & 0.575 \\
    \#\textbf{2} -- Text PT & {\color{cgood}\fatcheckmark} & {\color{cbad}\fatcross} & {\color{cbad}\fatcross} & 0.272 & \textbf{0.213} & \textbf{0.705} & \textbf{0.681} & \textbf{0.372} & \textbf{0.374} & \underline{0.604} \\
    \#\textbf{3} -- MIDI PT & {\color{cbad}\fatcross} & {\color{cgood}\fatcheckmark} & {\color{cbad}\fatcross} & 0.188 & 0.168 & 0.466 & 0.448 & 0.220 & 0.108 & 0.140 \\
    \#\textbf{4} -- Text\,\&\,MIDI PT & {\color{cgood}\fatcheckmark}$^{\ast}$ & {\color{cgood}\fatcheckmark} & {\color{cbad}\fatcross} & 0.265 & 0.206 & 0.620 & 0.586 & 0.328 & 0.315 & 0.573 \\
    \#\textbf{5} -- Text--MIDI SFT & {\color{cbad}\fatcross} & {\color{cbad}\fatcross} & {\color{cgood}\fatcheckmark} & \textbf{0.286} & 0.202 & 0.365 & 0.294 & 0.039 & 0.134 & \textbf{0.606} \\
    \hline
    \textbf{\textsc{Midi-LLM} (Ours)} & {\color{cgood}\fatcheckmark}$^{\ast}$ & {\color{cgood}\fatcheckmark} & {\color{cgood}\fatcheckmark} & \underline{0.282} & \underline{0.207} & \underline{0.626} & \underline{0.670} & \underline{0.335} & \underline{0.331} & 0.450 \\
    \hline
    \vspace{-1mm}
    \color{CadetBlue} Groundtruth Ref. &  &  &  & \color{CadetBlue}  \scriptsize 0.466 & \color{CadetBlue} \scriptsize 0.369 & \color{CadetBlue} \scriptsize 1.000 & \color{CadetBlue} \scriptsize 1.000  & \scriptsize \color{CadetBlue} 1.000 & \color{CadetBlue} \scriptsize 1.000 & \color{CadetBlue} \scriptsize 1.000 \\
    \bottomrule
\end{tabular}
\vspace{-3mm}
    \caption{Ablation study of \textsc{Midi-LLM} training stages, evaluated on TheoryTab lead sheets.
     Our full recipe achieves the best balance of music quality and textual control, validating the synergy of unimodal pre-training and multimodal SFT for specialized, data-constrained downstream tasks.
    {\small ($^{\ast}$: MusicPile text data (1.7B) included at continued pretraining with GigaMIDI.)}
    }\label{tab:ablations}
    \vspace{-3mm}
\end{table*}

\section{Experimental Setup}\label{sec:expr}
\subsection{\textit{TheoryTab} Lead Sheet Dataset}~\label{subsec:tt-ls-data}
We use the \textit{TheoryTab}~\cite{donahue2022melody} lead sheet dataset as a testbed for specializing \textsc{Midi-LLM} for music creative workflows.
A TheoryTab lead sheet consists of three tracks: \{Melody, Chord, Click\}.
The Click track marks beats and downbeats in time, which also imply the meter and tempo of the piece,
and the Melody and Chord tracks hold musical notes each specified by their (onset time, duration, pitch).
This format is conducive to training AMT-based~\citep{thickstun2024anticipatory} MIDI generation and infilling models, which have been proven helpful to songwriters in ideation and iteration~\citep{donahue2025hookpad}.

Besides musical content,
each lead sheet is annotated with metadata including its \textit{key}, \textit{genre}(s) (multi-label with 22 possible genres, e.g., pop, rock, country),\textit{ mood}(s) (multi-label with 42 possible moods, e.g., melancholic, playful, euphoric),
and five \textit{music-theoretical attributes}: \{chord difficulty (Ch-D), chord novelty (Ch-N), melody difficulty (Mel-D), chord-melody tension (Ch-Mel-T), bassline motion (Bass-M)\}
that are each binned into \textit{levels 1--7} evenly according to the raw score ranking among all lead sheets.

The lead sheets have an average length of 11.4 measures (std.~$=$ 6.1), or 165 notes (std.~$=$ 87), or 497 in MIDI tokens (std.~$=$ 261).
There are a total of 67K lead sheets in TheoryTab, amounting to
34M MIDI tokens,
which is $\approx$2\% the size of the LMD SFT dataset and
represents a more plausible data scale available for specialized music tasks.


\subsection{Model Implementation Details}

\subsubsection{CPT and Text-to-MIDI SFT}
We base our \textsc{Midi-LLM} implementation on Huggingface \texttt{transformers}~\citep{wolf2020transformers} and instantiate it
from the open-weight Llama 3.2 (1B)~\citep{grattafiori2024llama} checkpoint
with resized token embedding layer and output LM head to account for the expanded vocabulary (183K tokens in total).
Training is done
using \texttt{transformers}' trainer with FlashAttention-2~\citep{dao2024flashattention} and bf16 precision.
We perform 25K steps of gradient updates with AdamW~\citep{loshchilov2017decoupled} optimizer for both training stages.
We set the effective batch size to 512K tokens for CPT stage, 1M tokens for SFT stage, and learning rate to $2 \times10^{-4}$ with cosine decay.
The two-stage training run takes around 6 days on 4$\times$ H100 (SXM, 80G) GPUs.

\subsubsection{TheoryTab Lead Sheet Tasks and Finetuning}
We further finetune the model resulting from the two-stage training for TheoryTab lead sheet tasks.
Using the AMT framework~(cf.~Sec.~\ref{sec:token}),
we treat the tracks \{Melody, Chord, Click\} as three instruments \{Piano, Guitar, Drums\}.
The Click track is always given as anticipated controls,
and the model is tasked with generating Melody and/or Chord notes.
Following \citep{donahue2025hookpad}, we include four task modes \{From Scratch, Infill Melody, Infill Chords, Infill Both\}, where the right MIDI context for `Infill *' tasks is given as anticipated controls.
The task modes and target infill segments are sampled randomly and uniformly at finetuning.
As for text control, we organize all key, genre(s), mood(s), and the music-theoretical attribute levels, together with the task mode, into a JSON string, e.g., `{\small \textit{\{"task": "infill both", "key":"F major", "genre": "rock, metal", "mood": \dots (omitted)\}}}', and randomly drop each of them at finetuning (except task mode which is always included) for conditioning on any subset.
To enable classifier-free guidance~\citep{ho2022classifier} at inference, for 15\% of the time, only task mode is present in the JSON string.
Furthermore,
we augment the lead sheets on-the-fly using all 12 transpositions,
and time-stretched versions between 95\% to 105\% of the original tempo.

We use an effective batch size of 1M tokens, and a learning rate of $10^{-4}$ with 3K steps of cosine decay.
The model converges at about 1.5K steps, where we early-stop, taking around 1 day on 4$\times$ A6000 (48G) GPUs.

\subsubsection{Inference and Classifier-free Guidance}
We use \textit{nucleus sampling}~\citep{holtzman2020curious} with top $p = 0.98$ to balance musical diversity and coherence.
We apply serving-time optimizations in \texttt{vLLM}, including CUDA graphs, paged KV cache attention~\citep{kwon2023efficient},
and FP8-W8A8 dynamic quantization~\citep{shaw2024llm}.
In our pilot tests, \texttt{vLLM} sped up inference by 50\%+ compared to \texttt{transformers}' default API.

Also, in early MIDI infilling tests, we found that the surrounding MIDI context often dominates the text prompt, weakening the model's adherence to text-based controls.
To strengthen text control, we adopt classifier-free guidance (CFG)~\citep{ho2022classifier} following the LM-based setup in MusicGen~\citep{copet2023simple},
modifying the sampling distribution at each token step $t$ as:
\begin{equation}
\small
\begin{aligned}
\log & \, p_{\textrm{\textbf{CFG}}, \lambda} (\mathcal{X}_t \mid \bm{c}_{\textrm{M},t} \, , \bm{c}_{\textrm{T}}^{\textrm{cond}}) :=
\
\log p_{\theta}(\mathcal{X}_t \mid \bm{c}_{\textrm{M},t} \, , \bm{c}_{\textrm{T}}^{\textrm{base}}) \\
+ &\lambda \Bigl( \log p_{\theta}(\mathcal{X}_t \mid \bm{c}_{\textrm{M},t} \, , \bm{c}_{\textrm{T}}^{\textrm{cond}}) - \log p_{\theta}(\mathcal{X}_t \mid \bm{c}_{\textrm{M},t} \, , \bm{c}_{\textrm{T}}^{\textrm{base}}) \Bigr)
\end{aligned} \, ,
\end{equation}
where $\mathcal{X}_t$ denotes the set of candidate tokens,
$p_\theta(\cdot)$ is the model's estimated conditional distribution,
$\bm{c}_{\mathrm{M}, t}$ is the MIDI context that evolves with the generation,
$\bm{c}_{\textrm{T}}^{\textrm{cond}}$ is the intended text prompt,
$\bm{c}_{\textrm{T}}^{\textrm{base}}$ is the baseline text prompt specifying only task mode (e.g., `Infill Melody'),
and $\lambda$ is the CFG scale that allows users to modulate text control strength.


\begin{figure*}
  \centering
  \hspace{3mm}
  \includegraphics[width=0.3\textwidth,trim={0 3mm 0 4mm},clip]{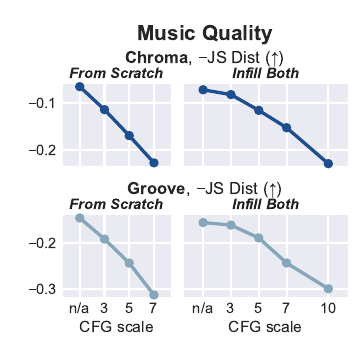}%
  \hfill
  \includegraphics[width=0.3\textwidth,trim={0 3mm 0 4mm},clip]{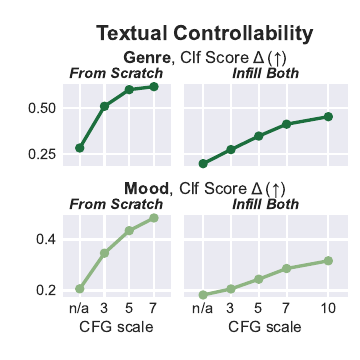}%
  \hfill
  \includegraphics[width=0.3\textwidth,trim={0 3mm 0 4mm},clip]{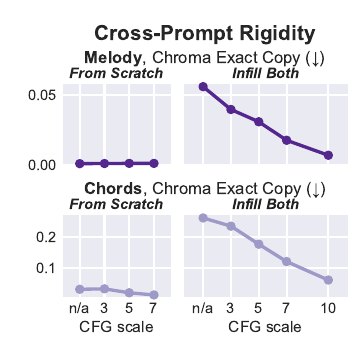}
    \hspace{3mm}
  \vspace{-4mm}
    \caption{Effects of classifier-free guidance (CFG) scale on lead sheet generation vs.~infilling.
    The quality-control tradeoff is markedly more \textit{inelastic} for infilling, requiring higher CFG scales to overcome conditioning from surrounding MIDI context.
    }
    \label{fig:cfg-tradeoff}
\vspace{-2mm}
\end{figure*}

\subsection{Evaluation Metrics, Protocols, and Baselines}
\subsubsection{On LMD Multitrack MIDIs}\label{subsubsec:eval-lmd}
Following the standard for text-to-music models~\citep{copet2023simple,evans2024fast,wu2024music},
we evaluate text-to-multitrack MIDI generations using the following two metrics computed with synthesized audio clips:
\begin{itemize}[nolistsep,itemsep=2pt,topsep=1pt,leftmargin=12pt]
    \item \textit{FAD}~\citep{kilgour2019fr}: measures the outputs' overall \textbf{quality} or realisticness using the Fr\'echet distance between the feature covariances induced by a set of model generations and those induced by a groundtruth set. 
    We employ VGGish~\citep{hershey2017cnn} as the audio feature extractor.
    \item \textit{CLAP}~\citep{wu2023large}: is meant to evaluate the model's \textbf{text control} adherence using pairwise feature cosine similarity between a contrastively trained text encoder (for text prompt) and audio encoder (for music output).
    We use the \texttt{music}
    checkpoint provided by LAION-AI.
\end{itemize}
We select a recent model, \textit{Text2midi}~\citep{bhandari2025text2midi}, trained also on MidiCaps\,+\,LMD
as our baseline.
Text2midi is based on an encoder-decoder architecture, with a (frozen) pretrained T5~\citep{raffel2020exploring} encoder providing text conditions, and a trained-from-scratch decoder modeling REMI-like~\citep{huang2020pop} music tokens.
Since our MidiCaps held-out split is different from that of Text2midi,
we identify the intersection (with 896 samples) for evaluation.

\subsubsection{On TheoryTab Lead Sheets}\label{subsubsec:tt-ls-protocol}

Among the 67K TheoryTab lead sheets,
we reserve 2K for validation LM loss monitoring,
and another 2K for testing.
From the final 2K, we construct two test sets: 
\begin{itemize}[nolistsep,itemsep=2pt,topsep=1pt,leftmargin=12pt]
    \item Test set \textbf{I -- Quality} focused: We give the Click track and the original key tag of all 2K lead sheets as conditions and have the model generate Melody and Chord tracks using the `From Scratch' task mode.
    \item Test set \textbf{II -- Control} focused: We select a subset of 250 lead sheets, and compose their key tag and Click track with \textit{single} controls including: 20 {\footnotesize(most common)} genres, 22 moods, and all levels 1--7 of the five music-theoretical attributes.
    We perform `From Scratch' generation, and in some experiments `Infill Both' on random 4-bar crops of each lead sheet, given the original MIDI context.
\end{itemize}
Instead of the system-to-system comparison done on MidiCaps\,+\,LMD (cf.~Sec.~\ref{subsubsec:eval-lmd}), we conduct training data ablations under an identical model architecture.
Specifically, we ablate one or more of the prior training stages: Llama text pretraining, our unimodal CPT, and text-to-MIDI SFT (see Table~\ref{tab:ablations} for details), and observe the performance impact on both test sets \textbf{I} and \textbf{II}.
Echoing the audio-domain FAD and CLAP metrics, here, we evaluate along quality and controllability axes based on symbolic music inputs:
\begin{itemize}[nolistsep,itemsep=2pt,topsep=1pt,leftmargin=12pt]
  \item \textbf{Music Quality:} Following~\citep{wu2020jazz}, we collect the bar-wise \textbf{chroma} (pitch) and \textbf{groove} (rhythm) vectors from all generations, and obtain histograms of chroma entropy and intra-piece Hamming groove similarity.
  We then compute the \textit{Jensen-Shannon (JS) distance} against corresponding histograms induced by groundtruth lead sheets.
  These metrics are bounded between $[0, 1]$, and lower$\,\downarrow$ values indicate a closer population-level pitch/rhythm feature behavior to human lead sheets, hence desirable.
  \item \textbf{Textual Controllability:} For \textbf{genre} and \textbf{mood}, we custom-train a joint multi-label classifier using the pretrained AMT-Large MIDI backbone~\cite{thickstun2024anticipatory}.
  We measure the JS distance between tag-wise classifier score distributions on generations given that tag (e.g., `{\small \textit{"genre":\,"rock"}}'), vs.~those given other tags (e.g., `{\small \textit{"genre":\,(some other tag)}}'), and macro-average over all possible tags (20 for genre; 22 for mood).
  We name this metric \textit{Clf.~Score $\mathbf{\Delta}\uparrow$}.
  For \textbf{music-theoretical attributes}, we follow~\citep{wu2023musemorphose} and compute the \textit{Spearman (ranking) correlation} $\rho\uparrow$ between the input control level (1--7) and raw attribute score computed from the generation.
\end{itemize}
In addition to data ablations, we use test set \textbf{II} to study the impact of varying CFG scales on `From Scratch' and `Infill Both' generations.
Besides quality and control, here, we also consider \textbf{Cross-Prompt Rigidity}.
Specifically, we compute the proportion of \textit{time-aligned} bars whose chroma vectors are an \textit{exact copy}\,$\downarrow$ under different text controls; higher values are undesirable as they suggest weaker influence from text control relative to that from MIDI context.

\


\vspace{-6mm}

\section{Quantitative Results and Discussion}\label{sec:quant-results}
Table~\ref{tab:vs-text2midi} summarizes system-to-system comparisons on text-to-multitrack MIDI generation.
Our Llama 3.2-based \textsc{Midi-LLM} outperforms Text2midi~\citep{bhandari2025text2midi} on both music quality and text control by considerable margins.
Also, thanks to \texttt{vLLM} integration, our model enjoys a 7--13$\times$ inference speedup despite its 5$\times$ model size, showing the advantage of leveraging optimizations readily available in the broader LLM ecosystem.
Though similar optimizations are possible for custom architectures like Text2midi,
they likely require substantial engineering efforts.
Moreover, using fp8 quantization yields an additional speedup ($\sim$20\%) and memory savings ($\sim$50\%) with a modest impact on quality.

Table~\ref{tab:ablations} shows the impact of \textsc{Midi-LLM} training stages on the downstream lead sheet generation task.
While Llama text pretraining (ablation \textbf{\#2}) alone establishes strong textual control,
it results in lower musical quality, underperforming the no-pretraining baseline (\textbf{\#1}) on multiple chroma and groove metrics.
Conversely, MIDI-centric settings (\textbf{\#3} and \textbf{\#5}) improve on music quality but exhibit poor or inconsistent controllability,
often regressing far below \textbf{\#1} on music-theoretical attribute controls.
Notably, the comparison between MIDI-only pretraining (\textbf{\#3}) and combined unimodal LLM \& MIDI pretraining (\textbf{\#4}) reveals that broad textual knowledge enhances not only textual control but also musical quality, suggesting a structural synergy between the two modalities.
This echoes, in an inverted sense, the findings in~\citep{papadimitriou2020learning},
where music pretraining was shown to help linguistic understanding.
Ultimately, our full \textsc{Midi-LLM} recipe achieves the most well-rounded performance with strong, balanced quality and text controllability.

Fig.~\ref{fig:cfg-tradeoff} illustrates the trade-off between music quality and text control adherence as a function of CFG scale.
In from-scratch generation, the model reaps strong controllability gains with a moderate CFG scale of 3.
However, the trade-off becomes much less elastic for infilling, requiring a high CFG scale of 7+ to obtain similar gains in control at a larger expense in quality. 
This behavior stems from the conditioning of surrounding MIDI context, as evidenced by the drastically high cross-prompt rigidity compared to unconstrained from-scratch scenarios (Fig.~\ref{fig:cfg-tradeoff}, right).
Without CFG, more than 25\% of time-aligned bar pairs share the exact \textit{same} chordal chroma profile under \textit{different} infilling text prompts.
The CFG scale thus serves as a mechanism to break this contextual rigidity, providing users the flexibility to navigate the tension between pre-existing musical material and their textual creative instructions.





\begin{figure}
\centering
\includegraphics[width=.92\columnwidth]{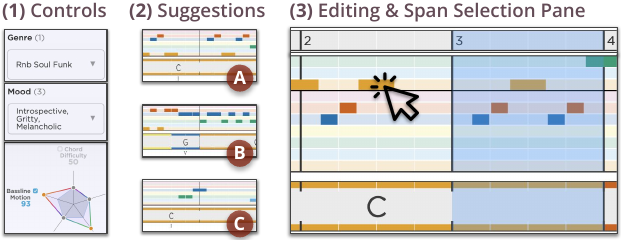}
    \vspace{-4mm}
    \caption{
    Co-creation user study interface components.
    }
    \label{fig:interface}
\end{figure}

\begin{figure}
\centering
\includegraphics[width=\columnwidth,trim={2.5mm, 2.5mm, 2.5mm, 2.5mm},clip]{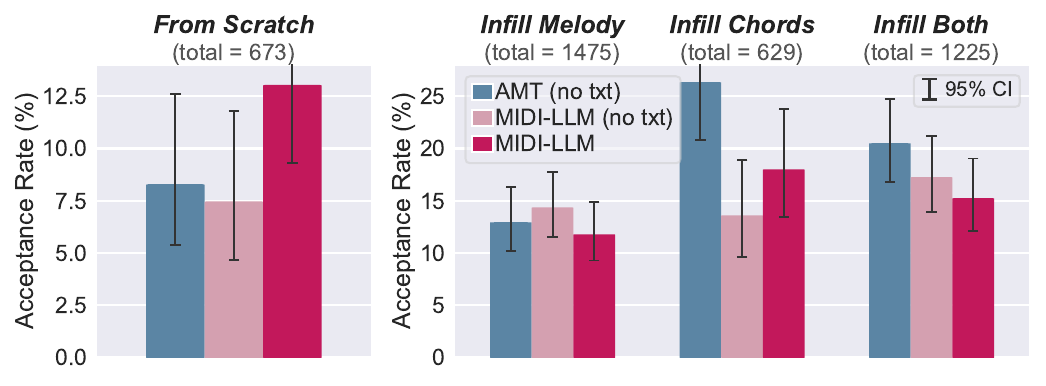}
    \vspace{-8mm}
    \caption{
    Co-creation user study acceptance results.
    }
    \label{fig:user-study}
\vspace{-3mm}
\end{figure}

\vspace{-3mm}
\section{In-the-Wild User Study}\label{sec:user-study}
To evaluate the utility
of \textsc{Midi-LLM} in real-world human-AI co-creation workflows,
we conduct an \textit{in-the-wild} user study via Hookpad, a lead sheet composition web platform
(Fig.~\ref{fig:interface}) developed by Hooktheory.
Users can specify any combination of the 7 controls (i.e., genre(s), mood(s), and 5 music-theoretical attributes),
and request model suggestions for \{From Scratch, Infill Melody, Infill Chords, Infill Both\} tasks.
A request is routed blindly to three backend models, all finetuned on TheoryTab:
\begin{itemize}[nolistsep,itemsep=1pt,leftmargin=12pt]
    \item \textbf{AMT (no txt)}~\citep{thickstun2024anticipatory}{\scriptsize  \,(360M)}: A MIDI-only model pretrained on~\citep{raffel2016lmd} that \textit{ignores} all of the 7 text-based controls.
    This model powered Hookpad Aria~\citep{donahue2025hookpad} prior to our study.
    \item \textsc{\textbf{Midi-LLM}} \textbf{(no txt)}: Our model with the 7 text-based controls \textit{stripped} from the prompt.
    \item \textsc{\textbf{Midi-LLM}}: Our full model with a fixed CFG scale\,$=$\,3.
\end{itemize}
Generated suggestions are returned in a randomized order.
Users can listen to the suggestions and decide which one to accept (or none) onto the lead sheet editing pane,
where they can edit the melody and chords, add their own musical material, and select spans for further model suggestions.

We deploy our user study to hobbyist songwriters 
who subscribe to Hookpad Aria,
which previously ran on the MIDI-only AMT model.
Over a 7-day study period, 58 unique users generated and listened to a total of 4,002 model suggestions and accepted 594 (14.8\% overall acceptance rate).
Fig.~\ref{fig:user-study} displays the acceptance rate comparisons faceted by the task. 
We highlight that, in the `From Scratch' task, our \textsc{Midi-LLM} achieves a 13.0\% acceptance rate, nearly double the 7--8\% from the two baselines (significant at  $p <0.1$).
This demonstrates that effective textual control is highly valued by users when establishing the foundations of a new lead sheet, enabling them to translate their creative intent directly into initial musical material.

In contrast,
the results for infilling are more nuanced, as the MIDI-only AMT baseline scores competitive or higher acceptance rates ($p < 0.05$ for `Infill Chords').
We propose a few possible explanations.
First, users may encounter a context-control mismatch, where specified textual controls are musically incompatible with the MIDI context, causing out-of-distribution scenarios.
Second, users may have a latent preference for musical continuity over textual control, favoring more often the suggestions that fit the MIDI context.
In this case, having an interface component (e.g., CFG scale) to balance the strength of text control and MIDI context could be helpful.
Finally,
there may be a familiarity bias towards AMT due to the users' prior experience.

After our study, we roll out MIDI-LLM broadly to all Hookpad Aria users, and observe a 22\% increase in accepted generations and infills, now totaling $\sim$1.5K per day.





\vspace{-2mm} 
\section{Conclusions and Future Work}
We introduced \textsc{Midi-LLM}, a recipe that improves text-to-MIDI generation by adapting LLMs.
Our quantitative experiments showcased the synergy between unimodal text/MIDI pretraining and paired text--MIDI supervision on a downstream data-constrained lead sheet task.
Furthermore, our in-the-wild user study demonstrated that text-based controls effectively aid real creators in zero-to-one musical ideation.

\textsc{Midi-LLM} opens promising avenues for future research.
For instance,
long-term deployment of co-creation platforms can facilitate user feedback collection for preference-tuning~\cite{ouyang2022training,rafailov2023direct}, establishing a continuous model improvement cycle.
Moreover, unifying symbolic and audio music modalities within the adaptation recipe can provide end-to-end support for the entire music composition and production pipeline.
These directions evolve \textsc{Midi-LLM} into a comprehensive partner for real-world music creative practices.


\bibliography{references}

%
%
%
%

\end{document}